# Triclinic nematic colloidal crystals from competing elastic and electrostatic interactions

Haridas Mundoor[1], Bohdan Senyuk[1], Ivan I. Smalyukh[1,2,3*]

[1]Department of Physics and Soft Materials Research Center, University of Colorado, Boulder, CO 80309, USA

[2]Department of Electrical, Computer, and Energy Engineering, Materials Science and Engineering Program, University of Colorado, Boulder, CO 80309, USA

[3]Renewable and Sustainable Energy Institute, National Renewable Energy Laboratory and University of Colorado, Boulder, CO 80309, USA

\* Correspondence to: ivan.smalyukh@colorado.edu

**Abstract:** Self-assembly of nanoparticles can enable composites with pre-designed properties but remains challenged by reproducing structural diversity of atomic and molecular crystals. We combine anisotropic elastic and weakly screened electrostatic interactions to guide both orientational and triclinic positional self-ordering of inorganic nanocrystals in a nematic fluid host. The lattice periodicity of these low-symmetry colloidal crystals is more than an order of magnitude larger than the nanoparticle size. Orientations of nanocrystals, as well as crystallographic axes of ensuing triclinic colloidal crystals, are coupled to the uniform alignment direction of the nematic host, which can be readily controlled on large scales. We probe colloidal pair and many-body interactions and demonstrate how triclinic crystals with orientational ordering of the semiconductor nanorods emerge from competing long-range elastic and electrostatic forces.

Since Einstein's seminal work on Brownian motion and Perrin's subsequent experiments (*1*) which showed that particles in colloidal dispersions obey the same statistical thermodynamics as atoms, the colloid-atom analogy has provided insights into physics of atomic systems through its application in studies exploring the dynamics of colloidal crystals and glasses (*2*). This analogy has inspired development of forms of self-assembly that attempts to reproduce the diversity of atomic crystals (*3*), although experimental realization of colloidal architectures with low symmetry, such as triclinic, remains challenging. At the same time, colloids are capable of designed control of their self-assembly, by exploiting aspects of particle shape (*4*) and topology (*5*), the dispersing medium's anisotropy (*6*) and composition (*7*), through DNA functionalization and origami-like designs (*8*), as well as the facile response of both particles and the medium to external fields (*9*, *10*). Long-range interactions attract a special interest as they can lead to sparse but ordered assembly of colloidal composites with unusual physical behavior (*9*). Long-range electrostatic repulsions in low-ionic-strength fluids (*10-13*) were used to obtain crystals and plastic crystals with high-symmetry colloidal lattices. These electrostatic interactions remain relatively isotropic at large separations despite anisotropic particles shapes (*10*). In nematic liquid crystal (NLC) hosts, highly anisotropic long-range colloidal interactions arise from the minimization of free energy associated with particle-induced elastic distortions of the NLC molecular alignment even when particles are spherical (*6, 14-16*), although the ability to control the shape of particles provides a means of guiding self-assembly (*17*). The diversity of elastic interactions, commonly resembling that of electrostatic dipolar and quadrupolar charge distributions (*6, 14, 17*), enabled colloidal self-assembly of lamellae and dipolar crystals (*18, 19*), albeit the inter-particle spacing in these lamellar and crystalline assemblies could be controlled only within a range comparable to the particle sizes.

We demonstrate that competition of long-range electrostatic and elastic interactions leads to a highly unusual self-organization of rod-like nanoparticles that exhibits both long-range orientational and triclinic positional ordering. Micrometer-range colloidal crystal lattice parameters of these assemblies, revealed by the three-dimensional (3D) optical imaging, are an order of magnitude larger than the size of constituent 30 nm × 150 nm colloidal semiconductor nanorods (fig. S1). We characterize pair interactions between the nanorods and both structure and dynamics of their dispersions at different surface charging and volume fractions. Although various dislocations, grain boundaries, vacancies and other defects are observed in these "soft" crystals, the crystallographic axes of triclinic lattices and colloidal nanorods tend to follow the direction of the alignment of the NLC host fluid. This preferred orientation of rod-like molecules is dubbed a "director" **n**, which can be controlled on large scales using approaches similar to those used to manufacture displays. This mechanical coupling is due to elastic free energy minimization at well-defined orientations of nanorods and triclinic colloidal crystals relative to a far-field director $\mathbf{n}_0$ and confining surfaces, potentially enabling device-scale self-assembly of tunable composites.

Semiconductor nanorods with the composition β-$NaY_{0.5}Gd_{0.3}Yb_{0.18}Er_{0.02}F_4$ engineered to define rod-like geometric shape and polarized up-conversion-based luminescence properties (Fig.1 and fig.S1) were synthesized using a hydrothermal method (*20-23*). Dispersion of nanorods in NLC (*9, 21*) was facilitated by surface functionalization of as-synthesized particles (*20-23*). In a typical process, 6mg of initially oleic-acid-capped nanorods in 8ml of cyclohexane were added to 4ml of DI water with small amount of hydrochloric acid to yield pH≈4 and stirred for 2h. During this process, oleic acid ligands became protonated and mixed with cyclohexane, leaving bare uncapped nanorods with positive surface charges (*21-23*). Nanorods were then washed with acetone 4-5 times, re-dispersed in water, and subsequently coated with methoxy-poly(ethylene glycol)silane (Si-PEG) (*21*). Typically, 5mg of Si-PEG dissolved in 1ml of ethanol was mixed with 5ml of nanorods dispersion in DI water (pH≈4) and stirred for 2h. After the reaction, particles were precipitated by centrifugation, dispersed in ethanol, and then re-dispersed in a pentylcyanobiphenyl (5CB) NLC via mixing and subsequent solvent evaporation at an elevated temperature of 70°C; the NLC composite was then quenched to room temperature while vigorously stirring (*9, 21*). NLC dispersions of nanorods were infiltrated into glass cells using capillary action. For planar boundary conditions for **n**, inner surfaces of cell substrates were coated with aqueous polyvinyl alcohol (1wt.% PVA) and unidirectionally rubbed. The cell gap thickness within 15–60µm was set using Mylar films. Surface charging of particles was characterized by probing their electrophoretic mobility within the aligned NLC cell under an electric field applied to in-plane electrodes (fig.S2). We controlled the effective surface charge per nanorod within $Z^*e \approx +(60-250)e$ (*22*), the Debye screening length in the nonpolar 5CB within $\xi_D = 0.1 - 1.2$ µm, and the nanorods surface potential (*13*) within $\Phi_0 = Z^* e \xi_D / [\varepsilon_0 \varepsilon (\Omega_{nr} + 4\pi l_{nr} \xi_D)] = 28\text{-}129\,\text{mV}$, where $Z^*$ is an effective number of elementary charges $e = 1.6 \times 10^{-19}$ C, $\Omega_{nr}$, $l_{nr}$, $\varepsilon$, $\varepsilon_0$ are a surface area and a length of nanorods, an average dielectric constant of 5CB and vacuum permittivity, respectively (*22*).

Dark-field and polarizing optical microscopies reveal positional ordering of nanorods (Fig.1A) and NLC alignment along $\mathbf{n_0}$ (fig.S3). The background-free 3D distribution of confocal luminescence from the nanorods, derived from an up-conversion process and collected while slowly scanning an infrared excitation laser beam (*22*), reveals colloidal crystals (Fig.1B,C,F). During this 3D imaging, the nanorods sense the potential landscape and jiggle around their minimum-energy triclinic lattice sites (Fig. 1B). To probe orientations of nanorods within the lattice, we measured luminescence intensity while rotating the analyzer with respect to $\mathbf{n_0}$ (Fig.1D). The intensity of emission at 552nm is maximum for an analyzer parallel to $\mathbf{n_0}$ at all studied concentrations, indicating that the nanorods align along $\mathbf{n_0}$ (fig.S1), as schematically depicted in Fig.1E. This nanorod orientation is consistent with minimization of total bulk elastic and surface anchoring free energy of NLC around PEG-functionalized nanoparticles (*9*). Weak quadrupolar particle-induced elastic distortions in the NLC bulk and the director orientation at nanorod surfaces compliant with tangential boundary conditions, except for the small surface point defect regions of discontinuity in the director field at the particle's poles dubbed "boojums". By combining the nanorod orientation and position data, we experimentally identify a primitive lattice cell based on confocal luminescence distributions due to eight representative nanorods (Fig. 1F) and then reconstruct a triclinic pinacoidal lattice (Fig.1G-I) with only the center inversion symmetry operation (*24*).

By probing nanorod displacements versus time with video microscopy in dark-field and up-conversion luminescence modes, we obtain the viscous drag (fig.S4) and the radial distribution function $g(r_{cc})$ in both nematic and elevated-temperature isotropic phases (Fig.2). Purely repulsive, direction-independent interactions are found in the isotropic phase of the nematic host at 40°C (Fig.2A). Once the host is cooled down to the nematic phase at 28°C, $g(r_{cc})$ reveals attractive forces in both dilute and concentrated dispersions (Fig.2,B and C). To probe the inter-nanorod pair interactions, we studied trajectories of two particles brought close to each other with optical tweezers and released (*22*). We used double helix point spread function (DHPSF) microscopy (*22, 25*) (Fig.3) to probe 3D positions of nanorods within the cell versus time with 7-10nm precision (*25*), as well as to characterize the corresponding orientations of their center-to-center separation vector $\mathbf{r}_{cc}$ relative to $\mathbf{n_0}$. The nanorods equilibrate at micrometer-scale pair separations and with $\mathbf{r}_{cc}$ tilting away from the sample plane while circumscribing a cone of q ≈ 49° ± 4° around $\mathbf{n_0}$; this would be impossible to quantify in 3D without DHPSF (Fig.3A-E). Pair interaction forces are highly anisotropic and long-ranged, with the angular distribution of $\mathbf{r}_{cc}$ orientations consistent with the cone of maximum-attraction angles expected for colloidal elastic quadrupoles (Fig.3) (*14-16*). Particle tracking yields the anisotropic distribution of nanorod displacements within the same intervals of elapsed time and the pair-interaction potential with a well-pronounced energetic minimum (Fig.3F). Since the attractive inter-nanorod van der Waals forces are negligible at the relevant $\mathbf{r}_{cc}$ (*2, 22, 26*), consistent with the fact that only repulsions persist when the host is heated to the isotropic phase (Fig.2A), we attribute this minimum in pair interaction energy to the competition of elastic and electrostatic interactions. Considering the classic use of multipolar expansions in both nematic elasticity and electrostatics (*1,2,14*), we use their leading terms (i.e., a monopole for electrostatics, because of the surface charging of nanorods, and a quadrupole for nematic elasticity, because of the director distortions shown in Fig. 1E) to model these interactions. The total potential can be approximately described as a superposition $u \approx u_C + u_{el}$ of quadrupolar elastic $u_{el}$ and electrostatic screened Coulomb $u_C$ interaction potentials (*2, 22, 27*), yielding $u(r_{cc}) = (A_1/r_{cc})\exp(-r_{cc}/\xi_D) + A_2(9 - 90\cos^2\theta + 105\cos^4\theta)r_{cc}^{-5}$. It fits well the distance dependence of the experimental potential (Fig.3F) and explains the equilibrium orientation of **r**$_{cc}$ at θ≈49°±4° to **n**$_0$, where $A_1$ and $A_2$ are fitting parameters dependent on the size and charging of nanorods, as well as alignment, elastic and dielectric properties (*22*). The Debye screening length $\xi_D$ = 0.34 µm derived from $A_1$ in $u(r_{cc})$ (*22*) matches $\xi_D$ independently obtained from fitting the repulsive electrostatic potential in the isotropic phase of 5CB, in which elastic forces vanish (inset of Fig.2A). The strength of the elastic quadrupole moment derived from $A_2$ as well as the material parameters such as elastic constants and anchoring coefficients are consistent with theories of quadrupolar elastic interactions and independent experimental measurements (*14-16, 22*).

As the concentration of nanorods increases, they exhibit gas-, liquid- and glass-like structural organizations, as well as crystalline order (Fig. 1 and figs.S5-S7) emerging at concentrations roughly consistent with the average equilibrium separations in colloidal pairs and lattices (Figs.1-3). Within crystallographic planes (100) (Figs.1 and 4A, table S1), **r**$_{cc}$ tends to align at angles θ≈49° to **n**$_0$, similar to pairs of nanorods. By analyzing particle displacement distributions associated with different lattice sites of the triclinic colloidal crystal (Fig.4B), we gain insights into the corresponding energy landscape (Fig.4C) and obtain the average spring constant of the crystal $k$≈0.63pNµm$^{-1}$. Translational order in our triclinic lattices (Fig.1A) is quantified by $g(r_{cc})$ (Fig.2C) and one-dimensional probability distribution function $g_{010}(r_{cc})$ (Fig.2D) calculated along the $a_2$ direction (Fig.1H,I). By probing the mean-square-displacement (MSD) $\langle\Delta r^2(t)\rangle$ of nanorods (Fig.4D, inset), we measure the so-called Lindemann parameter $\delta_L = [3\langle\Delta r^2(t\to\infty)\rangle/4r_{cc}^2]^{1/2}$ which is commonly used to characterize crystallization and melting transitions in terms of the MSD of particles around their ideal lattice positions as compared to their nearest-neighbor distance. By probing the Lindemann parameter versus the nanorod number density $\rho_N$ (Fig.4D), we find its concentration-dependent behavior and values corresponding to a crystallization transition consistent with that found in other condensed matter systems (*28*). Thermal expansion of ≈0.01°C$^{-1}$ of the lattice (fig.S8) stems from the decrease of an average NLC elastic constant $K$ and quadrupolar elastic forces with increasing temperature (by a factor of ≈3 when heated from room temperature to ≈34°C) (*22*).

Nanorods can be electrically concentrated and ordered starting from dilute initial dispersions (fig.S9), similar to crystallization of hard-sphere-like colloids subjected to electrophoretic or dielectrophoretic forces (*29*). The triclinic crystal order is facilitated by applying 300-900mV to transparent electrodes on inner substrates of the cell, lower than the threshold voltage needed for NLC switching. In response to these DC fields, the positively charged nanorods slowly move towards a negative electrode due to electrophoresis and eventually form a crystal as their concentration uniformly increases (fig.S9). These low voltages also facilitate uniform alignment of crystalline nuclei and healing of defects, as well as induce a giant electrostriction of the triclinic lattice, with ≈25% strain at fields of 0.03Vµm$^{-1}$ (fig.S8B). Finally, since NLC is switched at ≈1V (*9, 21*), colloidal crystal lattice orientations can be reconfigured while following the director rotation, albeit these processes are slow and complex. Electric fields, confinement in thin cells of thickness (≤15µm) incompatible with an integer number of primitive cells of the colloidal crystal, variations of nanorod concentration in the range exceeding that of an equilibrium triclinic lattice, as well as temperature changes control the primitive cell parameters (table S1) and prompt formation of defects ranging from edge dislocations (fig.S11) to vacancies and grain boundaries (*2, 22, 30*).

To conclude, we have introduced a highly tunable and reconfigurable colloidal system with competing long-range elastic and electrostatic interactions that lead to triclinic pinacoidal lattices of orientationally ordered nanorods. This unexpected triclinic crystallization of semiconductor particles at packing factors <<1% shows a potential for forming a large variety of mesostructured composites fabricated through self-assembly on device scales and tuned by weak external stimuli such as low-voltage fields and tiny temperature changes. The control of particle charging allowed for tuning triclinic lattice periodicity within 0.5-1.6μm, which can be extended further by tuning the strength of electrostatic interactions through doping or deionizing NLCs (*10-13*), as well as through using nematics with different properties. Since dipolar and other multipolar elastic colloidal interactions in NLCs can be introduced and guided by controlling boundary conditions at particle surfaces and since the control of NLC elastic constants may alter angular dependencies of these interactions (*22*), our studies set the stage for experimental and theoretical explorations of mesoscopic colloidal positional and orientational ordering that can enable engineering material properties through spontaneous ordering of nanoparticles.

**Acknowledgments:** This research was supported by the U.S. Department of Energy, Office of Basic Energy Sciences, Division of Materials Sciences and Engineering, under Award ER46921, contract DE-SC0010305 with the University of Colorado Boulder. We thank Paul Ackerman, Qingkun Liu, and Tom Lubensky for discussions.

**References and Notes:**

1. H. N. W. Lekkerkerker, R. Tuinier. *Colloids and the Depletion Interaction* (Springer Netherlands, 2011).
2. P. M. Chaikin, T. C. Lubensky, *Principles of Condensed Matter Physics* (Cambridge University Press, Cambridge, 1995).
3. V. N. Manoharan, Colloidal matter: Packing, geometry, and entropy. *Science* **349**, 1253751 (2015).
4. P. F. Damasceno, M. Engel, S. C. Glotzer, Predictive self-assembly of polyhedra into complex structures. *Science* **337**, 453-457 (2012).
5. B. Senyuk *et al*., Topological colloids. *Nature* **493**, 200-205 (2013).
6. P. Poulin, H. Stark, T. C. Lubensky, D. A. Weitz, Novel colloidal interactions in anisotropic fluids. *Science* **275**, 1770-1773 (1997).
7. S. Sacanna, W. T. M. Irvine, P. M. Chaikin, D. J. Pine, Lock and key colloids. *Nature* **464**, 575-578 (2010).
8. M. R. Jones *et al*., DNA-nanoparticle superlattices formed from anisotropic building blocks. *Nat. Mater.* **9**, 913-917 (2010).
9. Q. Liu, Y. Yuan, I. I. Smalyukh, Electrically and optically tunable plasmonic guest-host liquid crystals with long-range ordered nanoparticles. *Nano Lett*. **14**, 4071-4077 (2014).
10. B. Liu *et al*., Switching plastic crystals of colloidal rods with electric fields. *Nat. Commun.* **5**, 3092 (2014).
11. S. K. Sainis, J. W. Merrill, E. R. Dufresne, Electrostatic interactions of colloidal particles at vanishing ionic strength. *Langmuir* **24**, 13334-13337 (2008).

12. A. Yethiraj, A. van Blaaderen, A colloidal model system with an interaction tunable from hard sphere to soft and dipolar. *Nature* **421**, 513–517 (2003).
13. M. F. Hsu, E. R. Dufresne, D. A. Weitz, Charge stabilization in nonpolar solvents. *Langmuir* **21**, 4881–4887 (2005).
14. T. C. Lubensky, D. Pettey, N. Currier, H. Stark, Topological defects and interactions in nematic emulsions. *Phys. Rev. E* **57**, 610−625 (1998).
15. R. W. Ruhwandl, E. M. Terentjev, Long-range forces and aggregation of colloid particles in a nematic liquid crystal. *Phys. Rev. E* **55**, 2958-2961 (1997).
16. S. Ramaswamy, R. Nityananda, V. A. Raghunathan, J. Prost, Power-law forces between particles in a nematic. *Mol. Cryst. Liq. Cryst. Sci. Technol, Sect. A* **288**, 175-180 (1996).
17. C. P. Lapointe, T. G. Mason, I. I. Smalyukh, Shape-controlled colloidal interactions in nematic liquid crystals. *Science* **326**, 1083-1086 (2009).
18. J. C. Loudet, P. Barois, P. Poulin, Colloidal ordering from phase separation in a liquid-crystalline continuous phase. *Nature* **407**, 611-613 (2000).
19. A. Nych *et al*., Assembly and control of 3D nematic dipolar colloidal crystals. *Nat. Commun.* **4**, 1489 (2013).
20. F. Wang *et al*., Simultaneous phase and size control of upconversion nanocrystals through lanthanide doping. *Nature*, **463**, 1061-1065 (2010).
21. H. Mundoor, I. I. Smalyukh, Mesostructured composite materials with electrically tunable upconverting properties. *Small* **11**, 5572-5580 (2015).
22. Materials and methods are available as supplementary materials on *Science* Online.
23. N. Bogdan, F. Vetrone, G. A. Ozin, J. A. Capobianco, Synthesis of ligand-free colloidally stable water dispersible brightly luminescent lanthanide-doped upconverting nanoparticles. *Nano Lett.* **11**, 835-840 (2011).
24. D. E. Sands. *Introduction to Crystallography* (Dover Publications, 2012).
25. D. B. Conkey, R. P. Trivedi, S. R. P. Pavani, I. I. Smalyukh, R. Piestun, Three-dimensional parallel particle manipulation and tracking by integrating holographic optical tweezers and engineered point spread functions. *Opt. Express* **19**, 3835-3842 (2011).
26. C. A. S. Batista, R. G. Larson, N. A. Kotov, Nonadditivity of nanoparticle interactions. *Science* **350**, 1242477 (2015).
27. V. D. Nguyen, S. Faber, Z. Hu, G. H. Wegdam, P. Schall, Controlling colloidal phase transitions with critical Casimir forces. *Nat. Commun.* **4**, 1584 (2013).
28. R. W. Cahn, Materials science: Melting from within. *Nature* **413**, 582-583 (2001).
29. R. C. Hayward, D. A. Saville, I. A. Aksay, Electrophoretic assembly of colloidal crystals with optically tunable micropatterns. *Nature* **404**, 56–59 (2000).
30. I. I. Smalyukh, O. D. Lavrentovich, Anchoring-mediated interaction of edge dislocations with bounding surfaces in confined cholesteric liquid crystals. *Phys. Rev. Lett.* **90**, 085503 (2003).

**Figures**

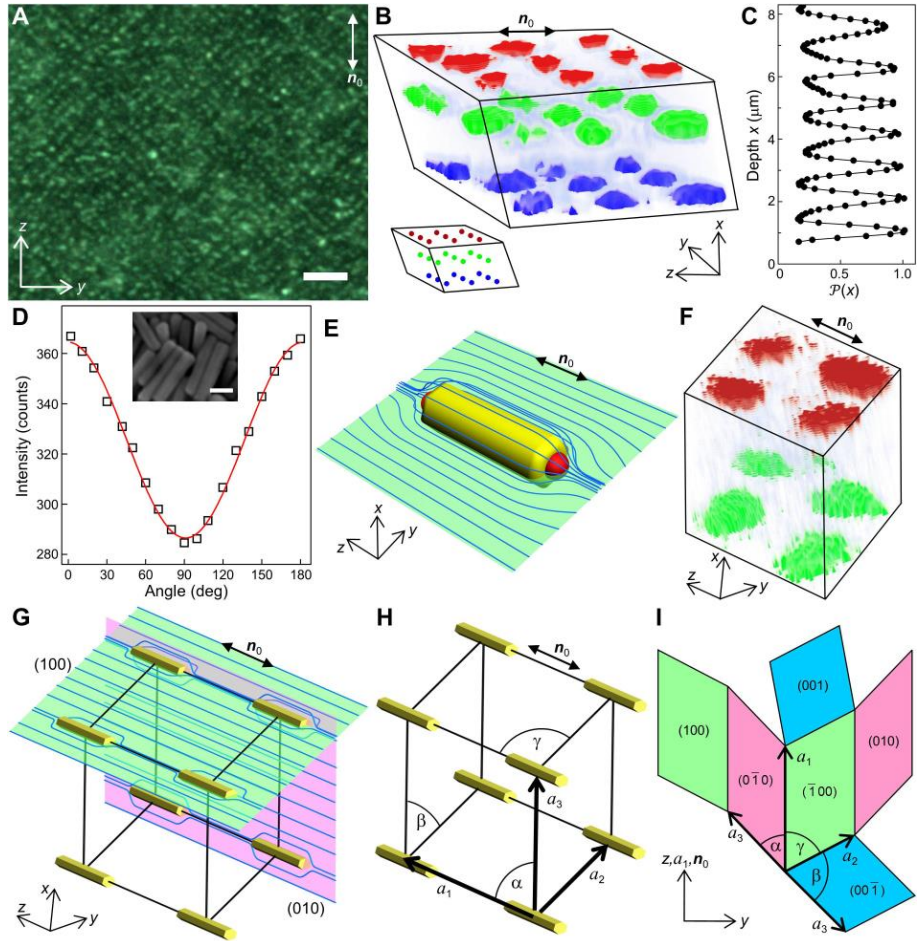

**Fig. 1. Triclinic crystal self-assembly of nanorods in NLC.** (**A**) Dark field micrograph of a crystal assembly with lattice parameters $a_1$=1.30±0.05 μm, $a_2$=1.01±0.05 μm; scale bar, 5 μm. (**B**) 3D micrograph showing luminescence from a triclinic colloidal crystal, which was reconstructed from slow confocal microscopy scanning (obtained within ≈3 min). It shows nanorods arrangements as they explore potential energy landscape near their minimum-energy lattice sites, with the luminescence signals from individual nanorods shown using red, green, and blue colors to identify location of particles in three consecutive planes parallel to confining glass substrates. The bottom inset shows center-of-mass positions of building blocks in a triclinic crystal with the same lattice. Lattice parameters based on averaging 18 independent local measurements are: $a_1$=1.49±0.06 μm, $a_2$=0.95±0.05 μm, $a_3$=1.20±0.05 μm, α=58°±2°, β=69°±2°. γ=49°±2°. (**C**) Probability distribution of finding nanorods at depth $x$ of the sample relative to the center of the first colloidal layer parallel to substrates (where $x$=0) calculated based on 3D luminescence imaging. (**D**) Luminescence intensity at 552nm versus analyzer rotation within 0-180°. Inset shows SEM image of the nanorods; scale bar, 50nm. (**E**) Schematic illustration of director distortions around a single nanorod, with the red hemispheres at the poles depicting two particle-induced boojums. These quadrupolar elastic distortions are axially symmetric with respect to the nanorod

axis parallel to **n**$_0$ and have mirror symmetry planes both parallel and orthogonal to **n**$_0$. (**F**) 3D micrograph showing a primitive unit cell of triclinic colloidal crystal, which was reconstructed from confocal scanning. (**G**-**I**) Schematics (not to scale) of primitive cell of a triclinic colloidal crystal, (G) showing local **n**(**r**)-distortions (blue lines) induced by nanorods, (H) defining parameters of a triclinic lattice, and (I) showing it unfolded..

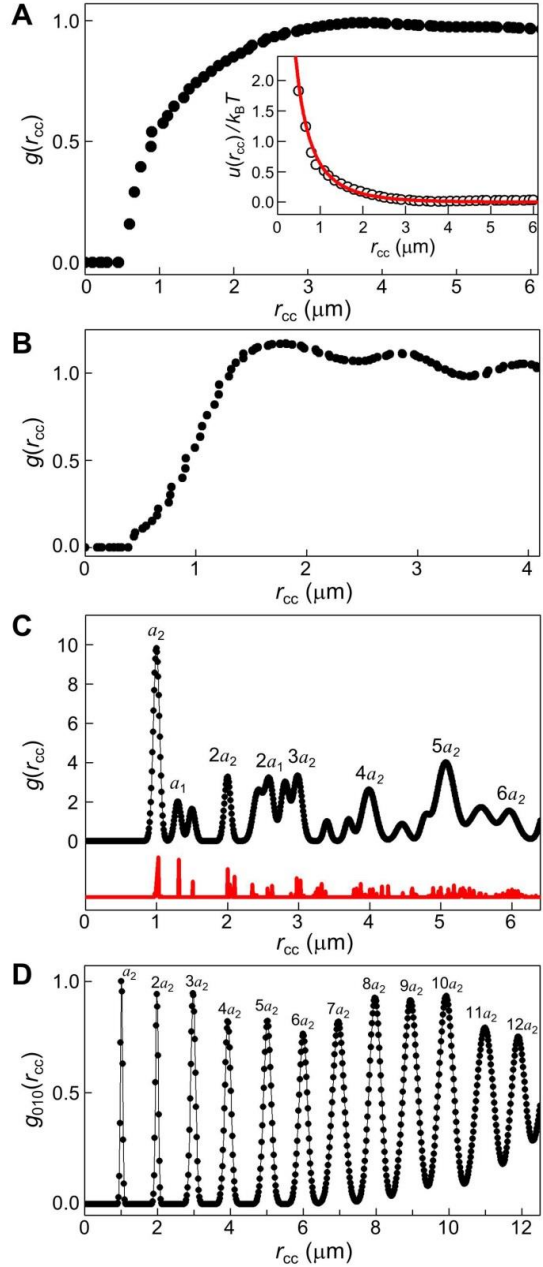

**Fig. 2. Ordering of nanorods in 5CB.** (**A**) Radial distribution function $g(r_{cc})$ for nanorods dispersion in isotropic phase, consistent with presence of long-range repulsive weakly screened electrostatic interactions. Inset shows pair potential extracted from these data and fit by a screened Coulomb interactions potential (*22*) ($k_B$, the Boltzmann constant $T$, an absolute temperature). (**B,C**) $g(r_{cc})$ for nanorod dispersion in nematic phase at (B) low $\rho_N \approx 0.35 \mu m^{-3}$ and (C) high $\rho_N \approx 4.5 \mu m^{-3}$ number densities of nanorods, with the emergent triclinic crystal ordering in the latter case.

Normalized red-colored $g(r_{cc})$ in (C) was calculated for the plane (100) of an ideal triclinic lattice with average dimensions determined from experiments, serving as an eye guide for the corresponding experimental peaks (●). (**D**) Probability distribution $g_{010}(r_{cc})$ calculated for the experimental triclinic lattice along $a_2$ by averaging data for [010] and [0$\bar{1}$0] directions. Numbers above the peaks indicate distances corresponding to integer numbers of lattice parameters $a_1$ and $a_2$.

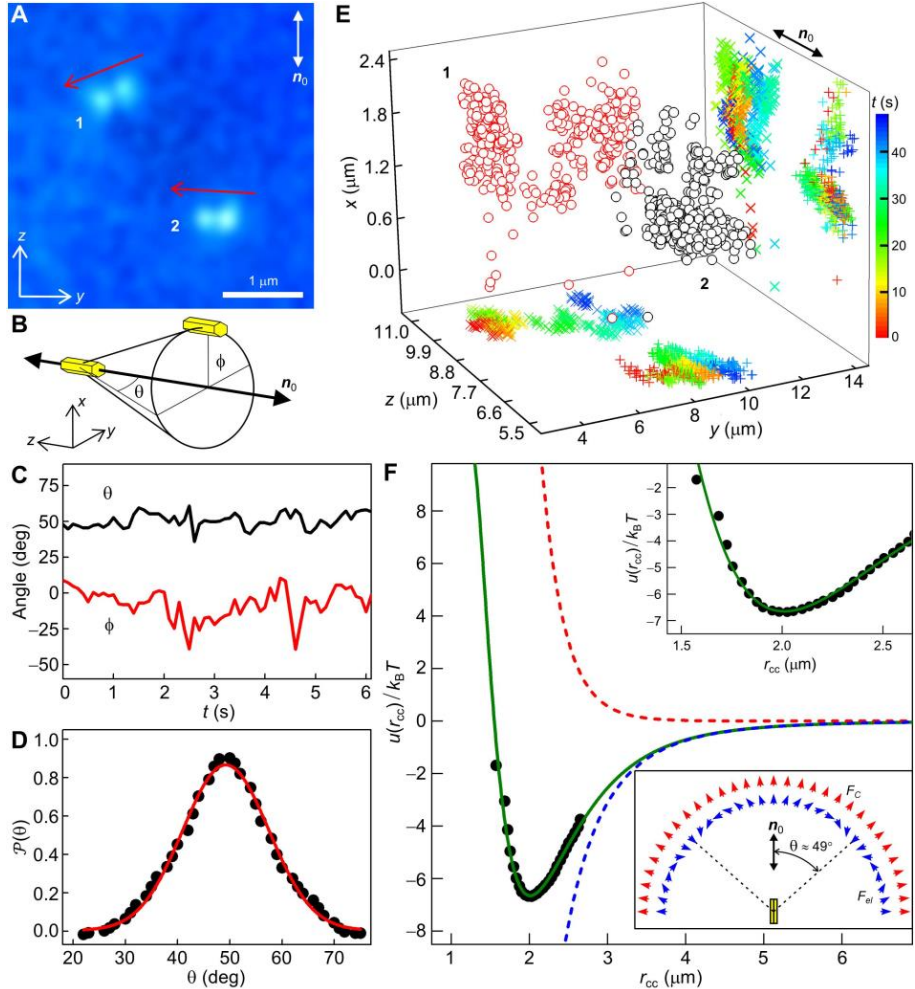

**Fig. 3. Pair interactions between nanorods.** (**A**) A representative DHPSF micrograph showing nanorods in 5CB at different depths of the cell corresponding to different orientation of bright-lobe pairs 1 and 2 (marked by red arrows) (*25*). (**B**) Schematic of two nanorods with $\mathbf{r}_{cc}$ tilted with respect to $\mathbf{n}_0$ and substrates, as characterized by angles θ and ϕ. (**C**) Typical changes of θ and ϕ over time. (**D**) Probability distribution (θ) of measured θ; a red line is a Gaussian fit. (**E**) 3D positions of two nanorods versus time *t* characterized with ≤10nm precision and depicted by red (1) and black (2) empty circles, as well as using their projections onto *zx*- and *zy*-planes, in which corresponding symbols (×) and (+) are colored according to the elapsed time scale. (**F**) Pair interaction potential extracted from experimental data and fit (green solid line) with the sum of screened electrostatic and elastic potentials. Dashed lines represent the electrostatic repulsive (red) and elastic attractive (blue) potentials. Top inset shows a zoomed view near the potential well

minimum. Bottom inset shows dependencies of directions of elastic force $F_{el}$ (blue arrows) and screened Coulomb electrostatic repulsion force $F_C$ on the angle between $\mathbf{r}_{cc}$ and $\mathbf{n}_0$.

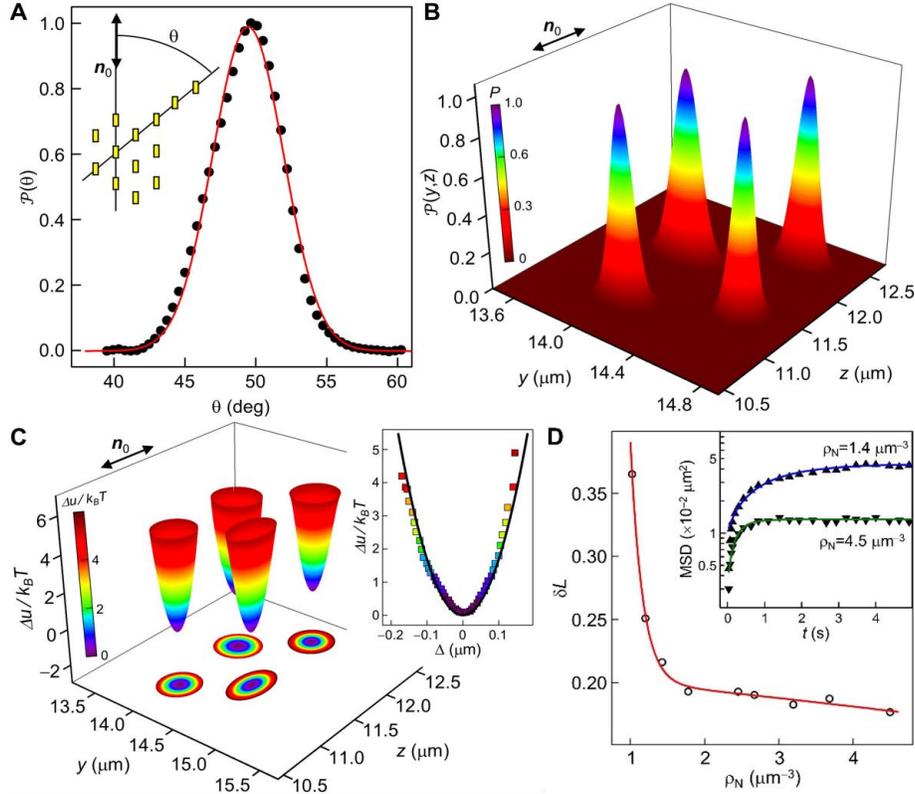

**Fig.4. Characterization of triclinic colloidal crystals.** (**A**) Probability distribution of an angle $\theta$ (inset) measured in the (100) plane of a colloidal crystal and its Gaussian fit (red line). (**B**) Probability $\mathcal{P}(y,z)$ distributions of positions within four lattice sites in the (100) plane of a triclinic crystal and (**C**) the corresponding potential landscape. Inset in (**C**) shows a local distance $\Delta$ dependence of relative potential energy experienced by nanorods. (**D**) Lindemann parameter $\delta L$ versus number density $\rho_N$ of nanorods. Inset shows MSD of nanorods versus time at $\rho_N$=1.4 µm$^{-3}$ (▲) and 4.5 µm$^{-3}$ (▼).

# Supplementary Materials

**Materials and Methods**

1. Semiconductor particle synthesis and initial dispersion

We synthesized lanthanide-doped β-NaY$_{0.5}$Gd$_{0.3}$Yb$_{0.18}$Er$_{0.02}$F$_4$ nanorods capped with an oleic acid (OA) by following a hydrothermal synthesis method reported previously (*20*), modified to achieve desired polarized luminescence properties and the rod-like geometric shape. The chemicals used for this synthesis, Ytterbium Chloride Hexahydrate (YbCl$_3$ 6H$_2$O), Yttrium Chloride Hexahydrate (YCl$_3$ 6H$_2$O), Erbium Chloride Hexahydrate (ErCl$_3$ 6H$_2$O), Thulium Chloride Hexahydrate (TmCl$_3$ 6H$_2$O), Gadolinium Chloride Hexahydrate (GdCl$_3$ 6H$_2$O), Ammonium Fluoride (NH$_4$F) and OA were purchased from Sigma Aldrich. Sodium Hydroxide (NaOH) was purchased from Alfa Aesar. Si-PEG (molecular weight 5000) was obtained from JenKem Technology. Briefly, 375 mg of NaOH was mixed with 1.875 ml of a deionized (DI) water, 6.25 ml of ethanol and 6.25 ml of OA. Additionally, 2.5 ml of 2 M solution of lanthanide chlorides with desired molar fraction, as needed for a specific doping concentration, was mixed with this solution. Finally, 1.25 ml of 2 M solution of NH$_4$F was added to the solution and stirred for 30 min to ensure proper mixing of the precursors. The final solution was transferred to a Teflon-lined autoclave (Hydrion Scientific, 25 ml), kept at an elevated temperature of 200ºC for 2 h and then allowed to cool down to the room temperature, completing the synthesis procedure. After the synthesis, nanorods were collected at the bottom of the reaction vessel, washed with ethanol and water (1:1) 4-5 times, and finally re-dispersed in cyclohexane. The particle size distribution was characterized based on electron microscopy images (inset of Fig. 1C and Fig. S1A). The particle composition described above was designed to yield polarized luminescence in the visible spectral range (Fig. S1B,C). This was achieved through the photon up-conversion processes (*21*) enabled by the ionic composition of nanorods and infrared excitation, allowing for a deep 3D imaging of positional ordering of nanorods within colloidal self-assemblies in nematic liquid crystal (NLC) (Fig. 1B).

2. Optical imaging using dark-field, luminescence, and confocal microscopies

We have utilized dark-field microscopy (Fig. 1A), luminescence microscopy (Fig. S2A), and up-conversion-luminescence-based confocal microscopy (Fig. 1B,C,F) techniques to image the structural colloidal organization in the NLC host. Planar alignment of the samples was verified with polarizing optical microscopy images of the sample under crossed polarizers (Fig. S3). The samples were illuminated with linearly polarized light, with the polarization defined by the polarizer *P*, and transmitted light from the samples was collected using a charge coupled device (CCD) camera (Flea-col, Point Grey or 18.2 Color Mosaic, SPOT insight). For the dark-field imaging (see, for example, Fig. 1A), which was enabled by a contrast of refractive index between nanoparticles and the surrounding nematic host fluid, we used a high magnification oil immersion objective (100×) with adjustable numerical aperture NA = 0.6−1.2, as well as a dark-field condenser U-DCW (NA = 1.2−1.4), both obtained from Olympus. In addition to imaging of concentrated dispersions of nanoparticles, this technique also allowed tracking the Brownian motion of individual colloidal objects and, thus, to probe viscous drag coefficients and diffusion properties (Fig. S4). The luminescence imaging was performed using a 100× oil immersion objective (NA = 1.4, from Olympus) and implemented using an inverted optical microscope (Olympus IX71). Nanorods were excited using the 980 nm output from a diode laser (Laserlands). To obtain a large-field illumination, we used a convex lens in order to produce a diverging laser beam before entering the objective. The images were also collected using a color CCD camera (Flea-col, Point Grey) with suitable optical filters added to an optical train to block the excitation beam. For imaging in isotropic phase of NLC, samples were heated by an objective heater (Bioptechs) mounted on the illumination objective. Figure S5 shows typical luminescence microscopy images of nanorods dispersed in NLC and corresponding to gas- (Fig. S5A), liquid- (Fig. S5B) and crystal-like structural colloidal organizations (Fig. S5C). This optical microscopy technique is also used as a means to characterize the local number density ($\rho_N$) of nanorods in the NLC host, as well as its changes in response to applied fields. These luminescence-based images complement our dark-field microscopy studies (Fig. 1A and Fig. S6) by providing details of structural organization within the typically smaller fields of view as compared to the ones in the dark-field images.

The 3D images of our colloidal crystals and other assemblies were obtained using a confocal microscopy setup built around an inverted microscope (Olympus IX81), albeit our implementation of depth-resolved confocal imaging reported in this work relies on the up-conversion-based luminescence rather than on fluorescence as in conventional confocal imaging. The pulsed output from a Ti:Saphire oscillator tuned to 980 nm was used to optically excite particles. The laser beam scans the sample to obtain depth resolved images in the following way. A set of galvano mirror controls a beam position in the *xy*-plane and a stepper motor controls the objective position along the *z*-direction parallel to a microscope's axis. The luminescence collected from the particles in epi-detection mode using a 100× oil immersion objective (NA = 1.4) was sent through a pinhole (which was confocal with the objective's focal plane) before being detected by a photomultiplier tube (PMT). This assured that only the luminescence from the focal plane of the objective was detected by PMT while the out-of-focal-plane light was effectively blocked, providing sub-micron diffraction-limited optical confocal resolution along the microscope's axis. Figure S7 as well as Fig. 1B,F show examples of the computer-reconstructed 3D images (presented using the ParaView software) of our NLC cells with colloidal dispersions at different volume fractions based on this type of confocal 3D imaging. Standard de-convolution and particle tracking procedures were applied to characterize the distribution of particle centers of mass to obtain data on particle distribution across the cell depth, such as the ones presented in Fig. 1C. Spectral luminescence measurements were performed with the same microscope setup while using a low-magnification objective (10×) to illuminate samples. The emission from the sample was collected by a 50× objective in forward detection mode and sent to a spectrometer (Ocean Optics, USB 2000) through an optical filter and an analyzer, yielding spectral and polarization dependencies of luminescence such as the ones shown in Fig. S1B,C and Fig. 1D.

3. Double-helix point spread function (DHPSF) imaging system

The conventional video microscopy limits video-rate tracking of spatial positions of colloidal particles with the high precision to the lateral directions in the plane orthogonal to an optical axis of a microscope. To overcome this limitation, we use optical imaging with a specially pre-engineered point spread function. The double helix point spread function (DHPSF) microscopy setup (*25*) is implemented using an inverted microscope Olympus IX71. The 980 nm output from a diode laser (Laserlands) is sent through a 100× oil immersion (NA = 1.4) objective to excite nanorods. The luminescence from particles, derived through a background-free photon up-conversion process, is collected using the same objective in epi-detection mode and projected on to a phase mask placed in the Fourier plane, using suitable lenses (*25*). The phase mask modifies the point spread function of single particles to yield two distinct lobes (Fig. 3A), which rotate with displacement of particles away from a focal plane. The image formed by the phase mask is projected to the electron multiplication charge coupled camera (EMCCD, from Andor Solis) for recording. The relative rotation angle of the bright lobes (Fig. 3A) is directly proportional to the displacement of particles along the microscope's axis. This provides the means of accurate tracking of vertical positions of nanorods with 7-10 nm precision based on the rotation of the bright lobes, such as the ones seen in Fig. 3A. Combined with the tracking of nanorod positions in the lateral plane of the microscope (*25*), this yields the full 3D video-rate particle tracking with 7-10 nm precision. Figure 3 shows how this DHPSF imaging enables accurate characterization of pair-interactions between nanorods in a uniformly aligned NLC host.

4. Control and analysis of surface charging

The surface charging of nanorods was tuned by controlled variation of the capping density of Si-PEG at the surface of particles as described below. The nanorods attain positive charges after removal of the OA molecules from their surfaces during the nanorod dispersion preparation, which is due to the protonation of particle surfaces in an acidic medium (*23*). To ensure a complete removal of OA molecules from the surface of particles, they were treated with pH ≈ 4 solution multiple times, as described in the main text. The Si-PEG ligands help to define tangentially degenerate surface boundary conditions for the NLC director and molecules at the nanorod surfaces and also tend to reduce the surface charging of nanorods. The electrostatic charging can be controlled by varying the density of Si-PEG chains attached at the surface of semiconductor particles. We have varied the Si-PEG concentration and reaction time to control the grafting density and thereby to tune the effective surface charging of the particles.

The total surface charge per particle prepared under different conditions was estimated from measurements of the speed of an individual nanorod moving in the NLC medium under an external electric field. For this experiment, planar NLC cells with in-plane electrodes separated by a distance ≈ 1 mm were used in a geometry depicted in Fig. S2A. Thin aluminum sheets (≈ 50 μm in thickness) were used as electrodes, serving also as spacers separating the confining glass plates and defining the NLC cell gap thickness. Copper wires were soldered onto the aluminum sheets and connected to an external power supply unit (DS340, Stanford Research Systems). A dilute spatially uniform dispersion of nanorods in NLC was infiltrated into the cell. While monitoring nanorod re-distribution using dark-field microscopy, the local concentration of the particles in NLC was tuned to approach a low number density per unit area ≈ 1000 mm$^{-2}$. The field-induced directional motion of the individual particles was observed with the help of the dark-field microscopy setup (Fig. S2A). When a DC voltage of 7 V was applied between the in-plane electrodes, the particle moved towards the negative electrode, consistent with the fact that the nanorods are positively charged. The motion of the particle was recorded using a CCD camera (Flea-col, Point Grey) and the corresponding speed of the particles was calculated from the extracted spatial displacements versus time (Fig. S2). In addition to the nematic phase of the fluid host of nanorods, the measurements were also repeated in an isotropic phase by increasing the temperature of this NLC host to 40°C, well above a nematic – isotropic phase transition temperature of the nanorod dispersion in 5CB ($\approx$34°C, about $\approx$1°C lower that the transition temperature of the pristine 5CB). This temperature control was achieved using an objective heater (Bioptechs). We also analyzed the motion of the particles when a low-frequency AC voltage (10 V at 1 Hz) was applied to the electrodes. In this case, the direction of particles motion oscillated with the periodic variation of the electric field polarity, as shown in Fig. S2D. The analysis of this periodic motion yields surface charging characteristics of nanorods consistent with that measured using a DC field described above.

The effective surface charge of the particle moving (note that the dynamics of this system is characterized by low Reynolds numbers) with a speed $v$ under the applied electric field $E = U/d$ was estimated from the balance of the Stokes viscous drag force $F_S = c_f v$ and the electric force $F_{el} = (Z^*e)E$ acting on nanorods, where $d$ is the distance between electrodes. This balance yields $Z^*e = c_f v d/U$, where $e = 1.6\times10^{-19}$ C is an elementary charge, $Z^*$ is an effective number of elementary charges on the nanorod's surface, and $U$ is an applied voltage. The friction coefficient $c_f$ was determined from the Einstein relation as $c_f = k_B T/D$, where $D$ is a particle diffusion constant at a temperature $T$. In order to estimate the diffusion constants from a separate experiment, the Brownian motion of a single particle in the NLC host was probed for prolonged times at different temperatures and $D$ was calculated from the distribution of displacements for the elapsed time interval $\Delta t$ (Fig. S4) extracted from the particle's trajectories, following the method described in details in Ref. 31. The diffusion of nanorods is anisotropic with respect to a far-field director $\mathbf{n}_0$ in a nematic phase, but direction-independent in the isotropic phase of the NLC host (Fig. S4).

We also independently estimated the Debye screening length by measuring the conductivity $\sigma$ before and after dispersing nanorods in the liquid crystalline host medium. We used glass substrates with ITO patterned electrodes (5 mm × 4 mm) to prepare a glass cell with spacing 10 µm. The conductivity of these samples was calculated by measuring a current flowing through the sample at a low-frequency voltage (10 V, 1-100 Hz) applied between electrodes, so that the capacitive resistance of NLC is negligible. The ionic strength of samples was calculated based on Walden's rule (*32*), and using the literature values of limiting equivalence conductance $\Lambda^E$ of hydrochloric acid (HCl) in ethanol (*33*) $\Lambda^E \eta^E = \Lambda^{LC}\eta^{LC}$, where $\eta^E$, $\eta^{LC}$ represents viscosities of ethanol and NLC, and $\Lambda^{LC}$ is an equivalence conductance of NLC. Following this, the Debye screening length was calculated as $\xi_D = (\varepsilon_0 \varepsilon k_B T/2N_a e^2 I)^{-1/2}$, where $N_a$ is the Avogadro's number, $\varepsilon$ and $\varepsilon_0$ are respectively an average dielectric constant of NLC (*37*) and vacuum permittivity and $I = \Lambda^{LC}/\sigma$ is the ionic strength/concentration. We obtained typical values of $\xi_D \approx 210$ nm for concentrated dispersions of nanorods in 5CB and $\xi_D \approx 378$ nm for the pure 5CB, consistent with the range of values obtained from fitting the colloidal interaction potentials (see the main text).

5. Characterization of colloidal dispersions based on the radial distribution function

To explore interactions and self-assembly of studied nanorods, we have used the two-dimensional radial distribution function $g(r) = \rho(r)/\rho_0$, where $\rho(r)$ is a density of pairs of particles separated by distance $r$ in a field of view or a video snapshot and $\rho_0$ is a density of uncorrelated particles in a homogeneous system (*34-36*). We used the dark-field video microscopy for data acquisition from diluted and concentrated samples in isotropic and nematic phases of the NLC host. In the low density regime [$\rho(r)\rightarrow0$], the pair distribution function is directly related to the effective pair interaction potential $u(r)$ as $g(r)=\exp[-u(r)/k_B T]$ (*35*), which allowed us to characterize the strength of electrostatic repulsions in dilute colloidal dispersions in the isotropic phase of the NLC host (Fig. 2A). The spatial positions of nanorods were tracked in the field of view of 47× 35 µm$^2$ using the tracking plugins of the ImageJ software (obtained from NIH). The number of analyzed frames for calculating $g(r)$ depended on the nanorod density in the NLC colloidal samples, which varied from less than 100 to about 2000 particles per frame in isotropic disordered and nematic ordered phases of the host fluid, respectively. Scattered experimental data were smoothed by using standard interpolation and averaging functions in OriginPro (OriginLab). For comparison purposes, we also computer-simulated images of nanorods ideally arranged into a triclinic lattice with parameters matching the experimentally measured ones. This allowed us to calculate $g(r)$ for a (100) plane of the triclinic crystal lattice to compare it with $g(r)$ extracted from the raw experimental data derived from dark-field videomicroscopy (Fig. 2C). The positions of experimental and simulated $g(r)$ peaks match well (Fig. 2C), especially in the near neighborhood of particles. Small differences at longer distances $r$ from the particles can be understood as arising from the effects of various dislocations, grain boundaries, and voids in the real experimental lattice, as well as from the limitations in our ability of tracking particles that exhibit substantial changes of intensity as their depth location varies with time. The latter source of artifacts arises from the fact that our triclinic colloidal crystal is intrinsically 3D in nature, so that the nanorods in their lattice sites can be occasionally missed during the videomicroscopy tracking because of the thermal fluctuations of their centers of mass positions along the depth of the NLC cell. Despite of all these challenges of experimental characterization, the agreement between experimental and calculated $g(r)$ for the (100) plane of our triclinic colloidal crystal is rather good (Fig. 2C). One-dimensional probability distribution function $g_{hkl}(r)$ (Fig. 2D) measures the probability of finding the center of mass of a nanorod at position $r$ with respect to a reference nanorod at $r=0$ in the direction defined by Miller indices [$hkl$] and was calculated for a structure corresponding to that observed in experiments (Fig. 1A) using video-tracking along the vector $a_2$ (including both [010] and [$0\bar{1}0$] directions) of the triclinic lattice (Fig. 2D). Confocal imaging allowed for similar characterization along the depth direction of the sample ($a_3$), which is shown in Fig. 1C.

6. Characterization of colloidal interaction pair-potential in the nematic phase of the NLC host

We have measured an interaction pair-potential between two electrostatically charged nanorods in a nematic phase of the NLC host through 3D tracking their positions by the DHPSF microscopy described above. The total interaction potential $u$ between two nanorods in NLC can be represented as the superposition of the pair potential $u_C$ due to screened Coulomb electrostatic repulsion forces, pair potential $u_{vdW}$ due to attractive van der Waals forces, and $u_{el}$ due to highly anisotropic quadrupolar NLC elastic interactions that can be either attractive or repulsive, depending on the orientation of the inter-particle separation vector $\mathbf{r}_{cc}$ relative to the far-field director $\mathbf{n}_0$. The total interaction potential reads:

$$u(r_{cc}) = u_C(r_{cc}) + u_{vdW}(r_{cc}) + u_{el}(r_{cc}), \text{ with}$$

$$u_C(r_{cc}) = A_1 \frac{\exp(-r_{cc}/\xi_D)}{r_{cc}}, \quad u_{vdW}(r_{cc}) = \frac{A_H}{6}\left(\frac{2l_{nr}^2}{r_{cc}^2 - 4l_{nr}^2} + \frac{2l_{nr}^2}{r_{cc}^2} + \ln\frac{r_{cc}^2 - 4l_{nr}^2}{r_{cc}^2}\right), \text{ and}$$

$$u_{el}(r_{cc}) = \frac{A_2}{r_{cc}^5}\left(9 - 90\cos^2\theta + 105\cos^4\theta\right),$$

where $\xi_D$ is the Debye screening length, $l_{nr}$ is a length of the nanorods, $A_H = 10^{-20}$-$10^{-19}$ is the Hamaker constant and $\theta$ is an angle between $\mathbf{r}_{cc}$ and $\mathbf{n}_0$. The first term $u_C(r_{cc})$ describing the electrostatic repulsion is in the screened Coulomb form (*38*) and the third term $u_{el}(r_{cc})$ describing the elasticity-mediated interaction potential is of quadrupolar type (*15, 16*), consistent with the configuration of the director distortions around nanorods with tangential surface anchoring (Fig. 1E,G). The van der Waals attraction $u_{vdW}(r_{cc})$ at the typical distances ($r_{cc}$ = 1-2 µm) measured for studied nanorods in the nematic host within colloidal pairs and lattices is many orders of magnitude smaller (constituting only $10^{-6}$ % of the total potential $u$) than the other two terms and, thus, can be neglected in the analysis of the total pair interaction potential $u$. The angular factor of the elastic term yields maximum strength of elastic attraction at $\theta \approx 49°\pm4°$, consistent with the experimental characterization of pair interactions and self-assembly (Figs. 1-4). The experimental fitting parameters $A_1 = \frac{(Z^*e)^2}{\varepsilon_0\varepsilon} \frac{\exp(l_{nr}/\xi_D)}{(1+l_{nr}/2\xi_D)^2}$ and $A_2 = \frac{8\pi W_p^2 l_{nr}^8}{90K}\left(1-\frac{W_p l_{nr}}{56K}\right)$, where $K$ is an average elastic constant of NLC and $W_p$ is a polar surface anchoring coefficient, match the ones estimated using the material parameters of our nanorod-NLC colloidal system measured independently for the same dispersion with $l_{nr} \approx 280$ nm, $Z^*e \approx +300e$, $\xi_D \approx 0.34$ µm, $\varepsilon \approx 11.1$, $\theta \approx 49°$, $K \approx 6\times10^{-12}$ N and $W_p \approx 10^{-4}$ J/m$^2$ (*15,37*). The value of the polar surface anchoring coefficient used in these estimates is consistent with the previous studies of rod-like nanoparticles functionalized with PEG (*9*). By using these parameters, one calculates the coefficients $A_1 \approx 4.06\times10^{-23}$ J m and $A_2 \approx 2.71\times10^{-48}$ J m$^5$, closely matching the fitting parameters $A_1 \approx 4.8\times10^{-23}$ J m and $A_2 \approx 2.7\times10^{-48}$ J m$^5$ obtained from analyzing the experimental data shown in Fig. 3. Small discrepancy can result from different size and effective charge of nanoparticles as well as from the complexity of surface anchoring effects. The elastic potential used in this analysis was adopted from the theory of quadrupolar elastic interactions described in Ref. 15, which deals with the finite anchoring regime and particle size being smaller than the anchoring extrapolation length $K/W_p$, albeit the theory was developed for spherical particles and its use to model our findings can be justified only at relatively large inter-particle distances relative to the particle size, as indeed is the case for our system.

7. Electric field control of colloidal self-assembly

When a concentration of nanorods in NLCs is increased gradually, the colloidal assemblies change accordingly (Figs. S6 and S7). However, the surface charging of nanorods and the NLC nature of the host medium enable facile electric control of colloidal assemblies (Figs. S8 and S9). The crystalline assembly of nanorods can be realized even when starting from a low-concentration initial dispersion by employing electrophoretic effects through applying a low-voltage electric field. To demonstrate this, the NLC cells with thickness $\approx 20$ µm were prepared using ITO-coated glass plates (Fig. S9A). After filling the cell with dilute dispersions of nanorods in NLC, a DC voltage of 300-900 mV was applied between the electrodes, which is smaller than a threshold voltage for the NLC realignment (~1 V for pure 5CB). Under the influence of this electric field, the positively charged nanorods moved due to electrophoresis and accumulated near the negative electrode, forming a triclinic crystalline assembly. Illustrating this capability of facile electrical control of self-assembly of nanorods, we show the dark-field micrographs before applying electric field (Fig. S9B) and the crystalline assembly mediated by an electric-field-induced local concentration of nanorods due to $U = 500$ mV voltage in the same sample area after 15 min (Fig. S9C). The electric field also induces the changes in the lattice parameters through electrostriction when such colloidal crystals self-assemble without or with the help of electrophoretic effects (Fig. S8B). For example, the crystalline colloidal organizations can be first realized by using electrophoretic effects and applied voltages of 300 mV, and then nanorods in colloidal lattices can be "squeezed" into assemblies with somewhat smaller lattice parameters at higher voltages (Fig. S8B). The latter effect manifests itself in the gradual decrease of the lattice parameters through a giant electrostriction of ≈ 25% at 0.03 V/µm (Fig. S8B).

8. Analysis of the packing fraction of nanorods in crystalline colloidal assemblies

To estimate the packing fraction, the volume of a primitive cell for a triclinic colloidal crystal system ($V_C$) was calculated using the measured values of lattice parameters (Table S1) and the formula $V_C = a_1\, a_2\, a_3\, (2\cos\alpha \cos\beta \cos\gamma - \cos^2\alpha - \cos^2\beta - \cos^2\gamma + 1)^{1/2}$. The volume occupied by the nanorods ($V_{nr}$) within the primitive cell is calculated by using the nanorod's geometric parameters of length and diameter measured based on transmission electron microscopy images (Fig. S1A). The measurements of lattice parameters were done based on dark-field optical micrographs (such as the one shown in Fig. 1A) and 3D confocal images (such as the one shown in Fig. 1B). The lateral spatial localization of centers of mass of nanorods based on scattering probed with dark-field microscopy is estimated to have precision in the range 7-10nm (*17*, *21*). The precision of determining the location of centers of mass of nanorods along the sample depth (in direction parallel to the microscope's axis) based on confocal luminescence images (*39*) is estimated to be ~50 nm. At the same time, our ability of accurately localizing the centers of lattice sites of the triclinic crystal lattice in the NLC host is additionally affected by optical effects, such as defocusing of light in the birefringent sample, by proximity of confining surfaces and defects like dislocations (Fig. S11), and also by thermal fluctuations of the particles within the potential energy landscape produced by the competition of elastic and electrostatic interactions. Taking these effects into consideration, when determining lattice parameters and packing fractions, we typically perform several local measurements to determine the primitive cell geometry and assess the accuracy of our measurements. The colloidal crystal packing fraction is then found as $f = V_{nr}/V_C$. Since the lattice parameter depends on the surface charging of nanorods, calculated values of packing fraction for crystalline assemblies under different conditions range within $f$=0.008-0.064%. These values are much smaller than what was previously achieved for other low-symmetry colloidal crystals. This finding is natural because the dimensions of nanorods are much smaller as compared to the crystallographic primitive cell dimensions, as determined by the competition of the long-range electrostatic and NLC-mediated elastic colloidal interaction forces.

9. Control of crystal lattices by tuning material and confinement parameters

The crystalline assembly of nanorods in NLC is mediated by the competing actions of electrostatic repulsion and anisotropic elastic forces between the particles. The quadrupolar type elastic interaction potential prompts orientation of $\mathbf{r}_{cc}$ at an angle $\theta \approx 49°$ with the NLC director $\mathbf{n}_0$, as described in details above and in the main text. Although the crystalline symmetries of nanorods in NLC are affected by confinement, geometric cell parameters, gravity, applied electric fields, and other factors, the self-assembly of nanorods is largely determined by the equilibrium angle $\theta$ that in the studied sparse triclinic crystalline assemblies is found to be close to that exhibited by pair interactions (Figs. 3 and 4). This leads to the triclinic pinacoidal colloidal crystal lattice found in our experiments (Fig. 1). Although the NLC host has a uniaxial nonpolar symmetry invariant with respect to rotations around the far-field director $\mathbf{n}_0$ (the $D_{\infty h}$ group), so that, in principle, many different orientations of the triclinic lattice with respect to $\mathbf{n}_0$ could be possible, the interaction of nanorods in different crystallographic planes with the confining substrates lifts this degeneracy. Figure S10A-D shows schematics of a triclinic primitive cell for two different orientations of the triclinic crystallographic planes with respect to substrates, which we observed in our samples. Table S1, in addition to the parameters of the most common triclinic lattice with the orientation of the crystallographic plane (100) parallel to the glass substrates (Fig. 1 and Fig. S10A,C), also provides the primitive cell parameters for the triclinic cell oriented with the plane (010) parallel to the glass substrates (Fig. S10B,D). The fact that the ratios between the different magnitudes of vectors of the primitive lattice are different for the two orientations is natural and can be attributed to the effects of the cell confinement. Indeed, the proximity of the glass substrates locally lowers the $D_{\infty h}$ symmetry of the ground-state NLC, making the energetic costs of deformations of the director field in the plane of the cell different from the deformations across the cell thickness, which subsequently leads to the symmetry breaking and different ratios of lattice parameters for the two orientations of the triclinic lattice (Table S1 and Fig. S10A-D). In addition to the two observed orientations of the crystallographic planes with respect to confining cell substrates (Fig. S10A-D), we also found regions with opposite tilt of the (001) and ($00\bar{1}$) planes as well as the (010) and ($0\bar{1}0$) planes, which is consistent with the nonpolar symmetry of the director of the NLC host fluid. This alternation of tilting direction of the crystallographic (001) and ($00\bar{1}$) planes as well as the (010) and ($0\bar{1}0$) planes also contributes to the large number of defects, such as the tilt grain boundaries and dislocations (Fig. S11) observed in our colloidal crystals.

    Although the interaction of self-assembled colloidal crystals with confining substrates and the fine dependence of lattice parameters on material and geometric cell parameters fairly complicate colloidal self-assembly of nanorods in the NLC fluid, these factors also allow for potential means of controlling this self-assembly. In principle, the symmetry of crystals formed by nanorods can be tuned by modifying surface boundary conditions on nanorod surfaces (e.g. changing them from tangential to homeotropic or conically degenerate), by using materials with different anisotropies and strengths of elastic constants (for example, thermotropic small molecule and polymeric NLCs as well as chromonic lyotropic NLCs are known to all have dramatically different elastic constant anisotropies), by inducing pretilt of the far-field director of the NLC host with respect to the confining cell substrates, which may lift the degeneracy of tilting of the (001) and ($00\bar{1}$) planes as well as the (010) and ($0\bar{1}0$) planes, and so on. These different conditions are expected favoring different equilibrium angles θ. For illustration purposes, we show examples of crystal lattice with θ = 45° and 60° (Fig. S10E,F). When the effects of confinement are neglected and θ = 45°, nanorods could form a body centered tetragonal lattice with lattice parameters $a_1=a_2 \neq a_3$ and α=β=γ=90° (Fig. S10E). In a similar case for θ = 60°, the expected crystal symmetry is trigonal rhombohedral (Fig. S10F), with lattice parameters $a_1=a_2=a_3$ and 120°>α=β=γ≠90°. A non-exhaustive preliminary exploration of these tuning possibilities applied to our triclinic colloidal crystals already allowed for a rather significant control of lattice parameter ratios $a_2/a_1$ within 0.6-0.8 and $a_3/a_1$ within 0.6-0.9. The parameter $a_1$ was controlled within 650-1600 nm by adjusting the surface charging. These are only some of many examples of the possibilities that can arise in the case of colloids with quadrupolar symmetry of elastic distortions. Our approach can be also extended to elastic dipoles and other multipoles, as well as to the particles with shapes beyond that of cylinders and spheres, where geometric shapes could be further used to guide colloidal self-assembly (*17*).

**Supplementary Figures**

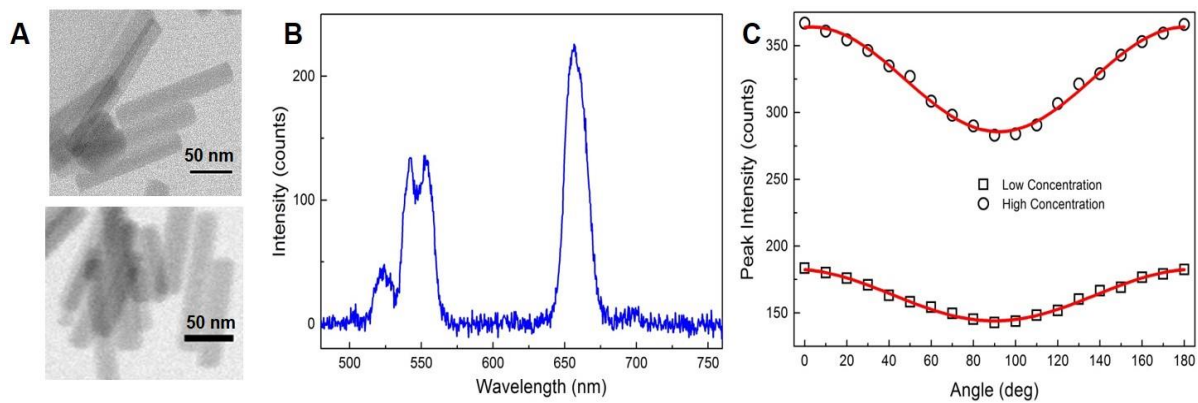

**Fig. S1. Nanorods with polarized luminescence engineered to probe orientational and positional ordering in colloidal assemblies.** (A) Transmission electron microscopy images of nanorods. (B) Emission spectra measured for nanorods on a glass substrate, when excited using 980 nm laser light. (C) Angular variation of nanorod's polarized emission intensity in the NLC dispersions measured at 552 nm when the analyzer $A$ was rotated within 0-180° with respect to $\mathbf{n}_0$, for low (volume fraction ≈ $5.3\times10^{-5}$, $\rho_N$ ≈ 0.5 µm$^{-3}$) (□) and high (volume fraction ≈ $3.2\times10^{-4}$, $\rho_N$ ≈ 3 µm$^{-3}$) (○) concentrations of nanorods in 5CB.

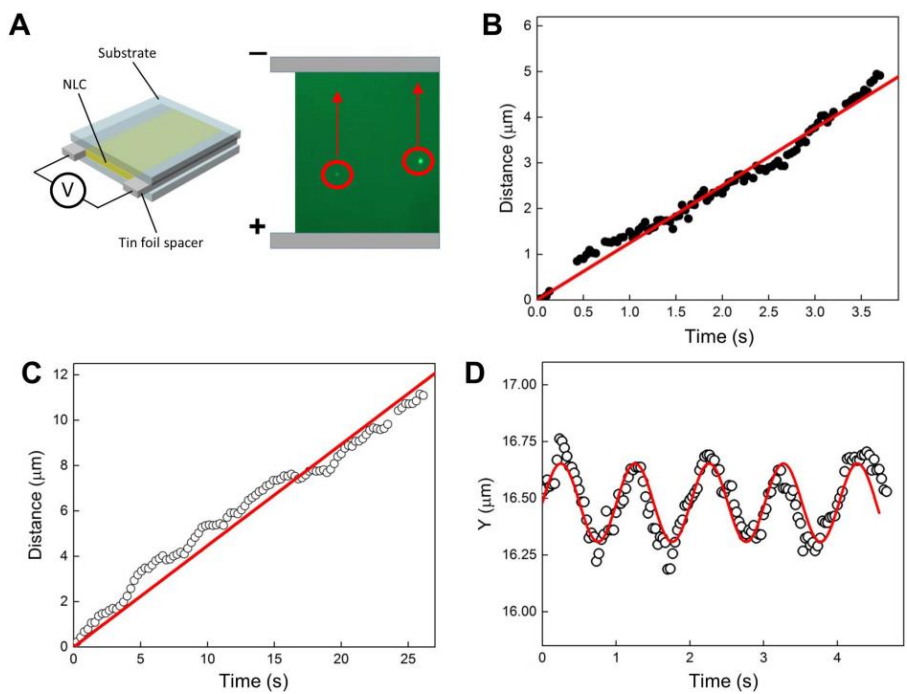

**Fig. S2. Characterization of surface charging of nanorods.** (A) Schematic representation of a NLC cell construction used for characterization of nanorods charge. The dark-field image shown

on the right depicts two nanorods in NLC. Motion of these nanorods was used for charge estimation. The direction of motion of nanorods is indicated by the arrows and the polarity of the DC applied field is marked on the electrodes. (B-D) Plots of displacement of particles with elapsed time when electric field is applied between the in-plane electrodes for (B) a DC voltage $U = 10$ V in the NLC's isotropic phase, (C) DC voltage $U = 7$ V in the nematic phase at room temperature and (D) for the case of applied low-frequency AC voltage $U = 10$ V (at 1Hz) in the nematic phase.

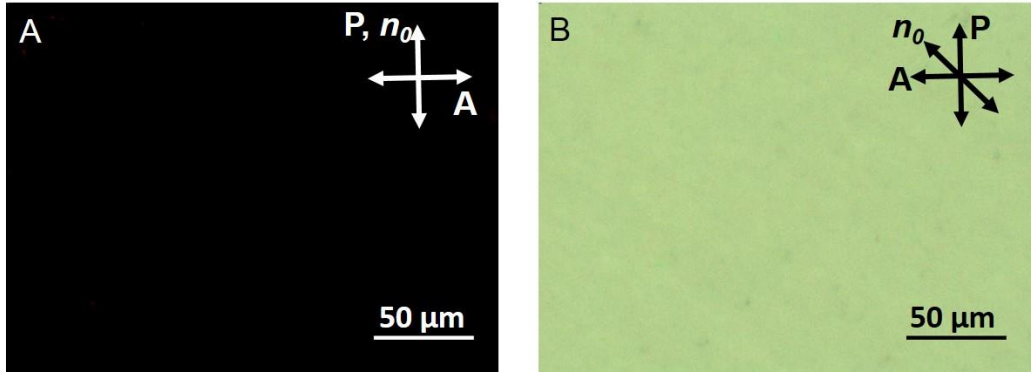

**Fig. S3. Background orientational ordering of the NLC host.** Polarizing optical micrographs showing the nanorod-NLC composite between cross polarizers when the far-field director $\mathbf{n}_0$ is kept (A) at 0° and (B) 45° with respect to the polarizer $P$, revealing the overall uniform planar alignment of the sample.

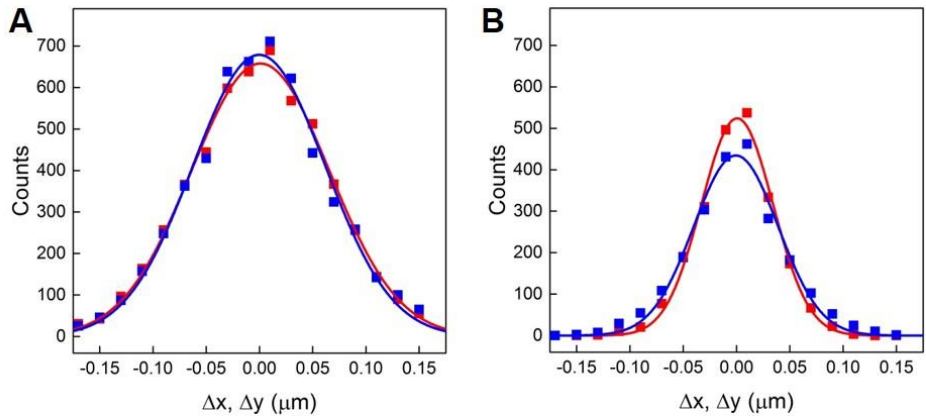

**Fig. S4. Self-diffusion of individual nanorods in dilute dispersions in the NLC.** (A,B) The distribution of the displacement of nanorod in the lateral $x$- and $y$-directions characterized for the elapsed time interval $\Delta t = 66$ ms of the Brownian motion in (A) isotropic and (B) nematic phase of the NLC host. The far-field director $\mathbf{n}_0$ in (B) was pointing in the $y$-direction.

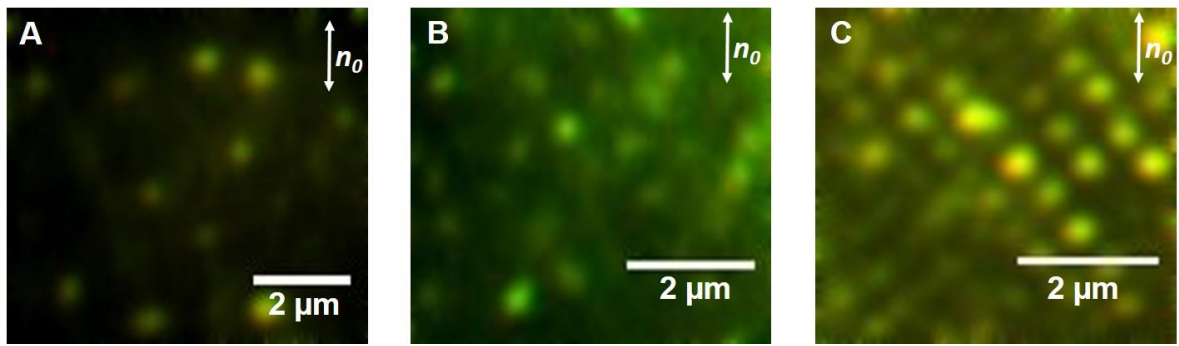

**Fig. S5. Nanorod dispersions in 5CB at different volume fractions and structural organizations.** (A-C) Luminescence microscopy images of particles dispersed in NLC as the concentration of nanorods is increased gradually, yielding colloidal analogs of (A) gas at volume fraction $1.7 \times 10^{-5}$ ($\rho_N = 0.16$ µm$^{-3}$) (B) liquid at volume fraction $5.7 \times 10^{-5}$ ($\rho_N = 0.54$ µm$^{-3}$) and (C) crystalline organization at volume fraction $4.7 \times 10^{-4}$ ($\rho_N = 4.5$ µm$^{-3}$).

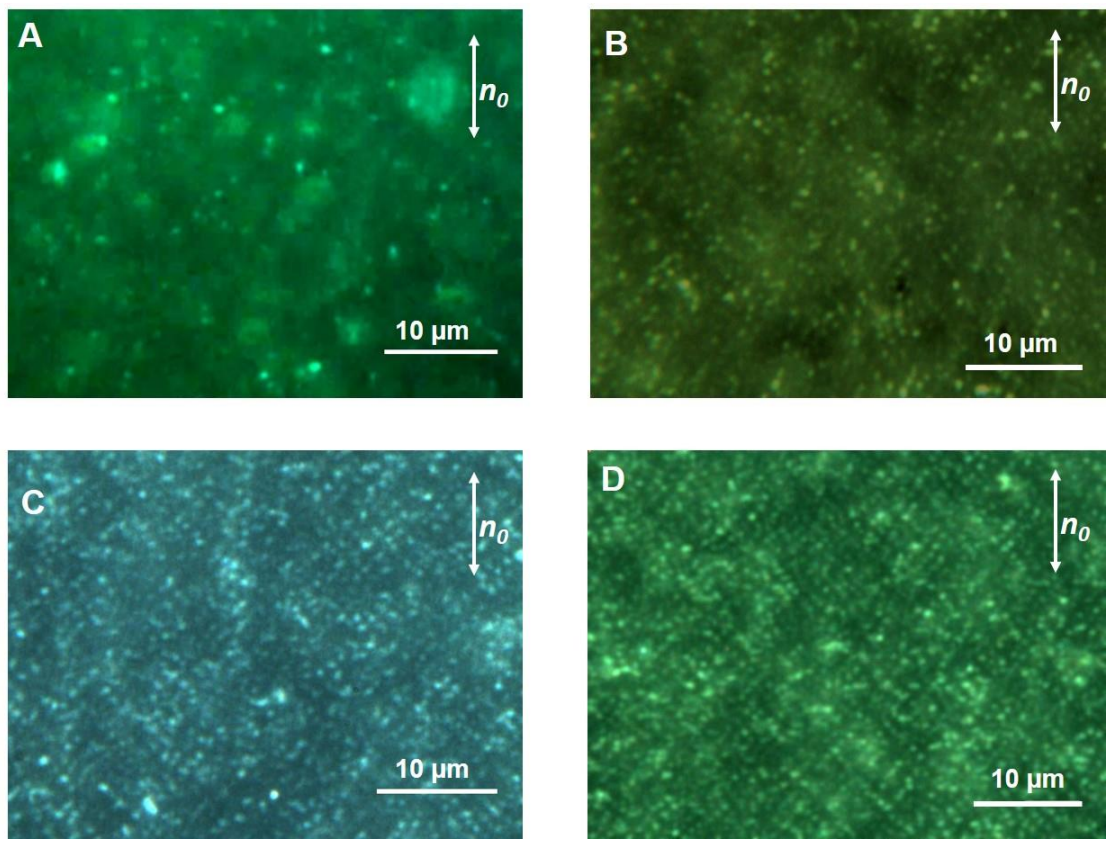

**Fig. S6. Colloidal organization of nanorods in 5CB versus concentration.** (A-D) Large-area dark-field microscopy images of nanorods dispersed in NLC when their concentration is increased gradually, yielding colloidal dispersions that resemble (A) gas- (B) liquid- (C) glass- and (D) crystal-like structures.

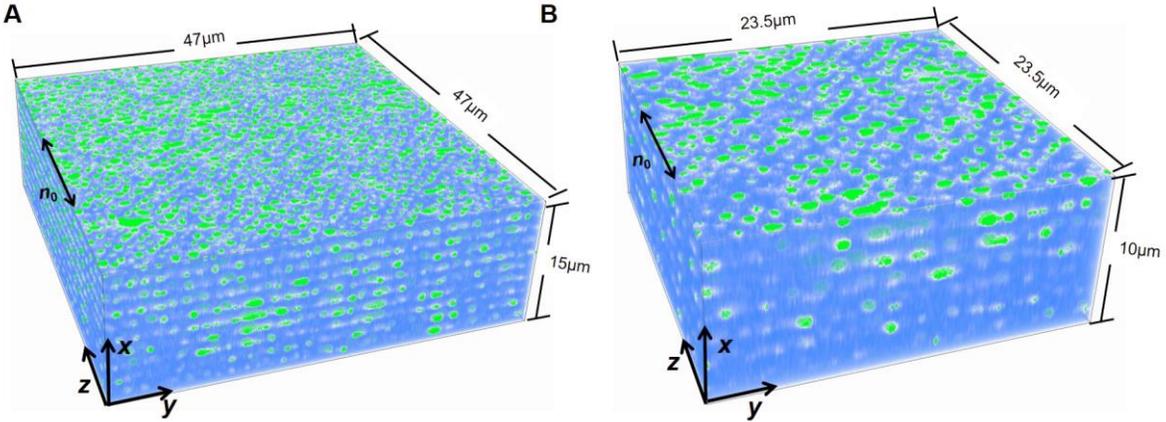

**Fig. S7. Examples of 3D confocal luminescence images of nanorod dispersions in NLC.** (A,B) 3D images of experimental cells for (A) zoomed-out and (B) zoomed-in views of the colloidal assembly.

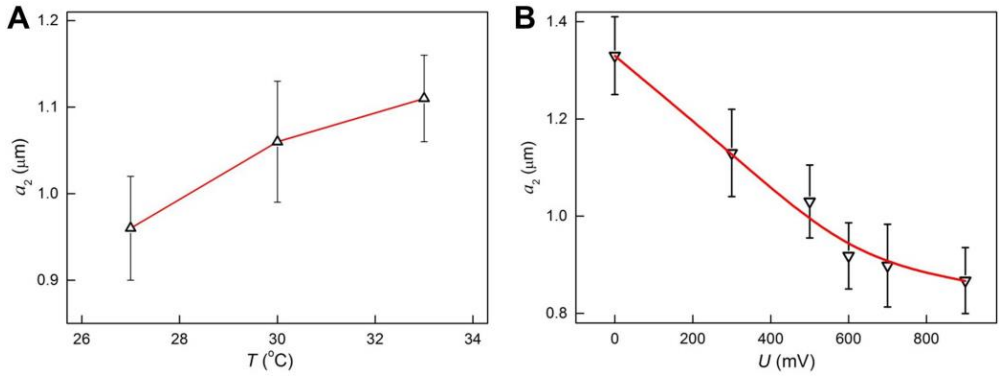

**Fig. S8. Thermal expansion and electrostriction of triclinic colloidal crystals.** The effect is characterized by probing changes of the lattice constant $a_2$, defined in the main-text Fig. 1D. (A) Lattice expansion prompted by the temperature increase from 27°C to 33°C. (B) Electrostriction of the triclinic crystal lattice prompted by a DC voltage change within 300 – 900 mV.

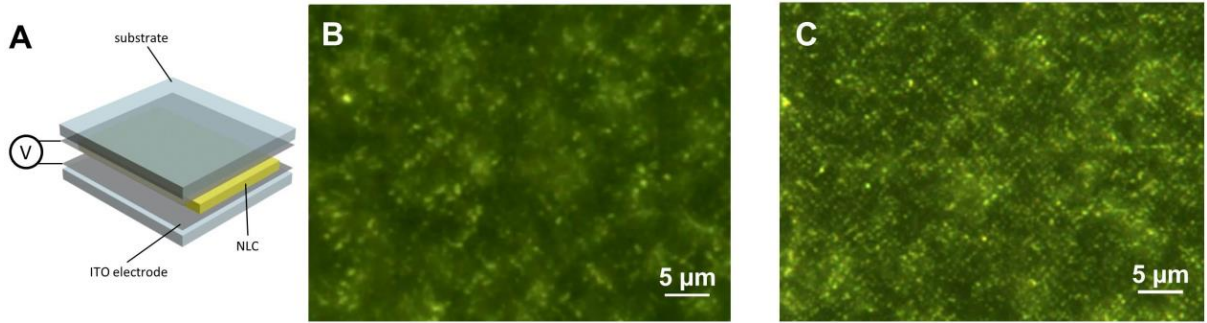

**Fig. S9. Electric control of colloidal assembly of nanorods in NLC.** (A) Schematic representation of the NLC cell showing arrangements of the electrodes used for studying the applied field effect on crystal formation and alignment. Dark field microscopy image of nanorods in NLC at a low concentration (volume fraction ≈ 3.2 ×$10^{-5}$, $\rho_N$ ≈ 0.3 µm$^{-3}$) (B) before applying the field and (C) the crystalline arrangement formed after applying 500 mV DC voltage between the electrodes.

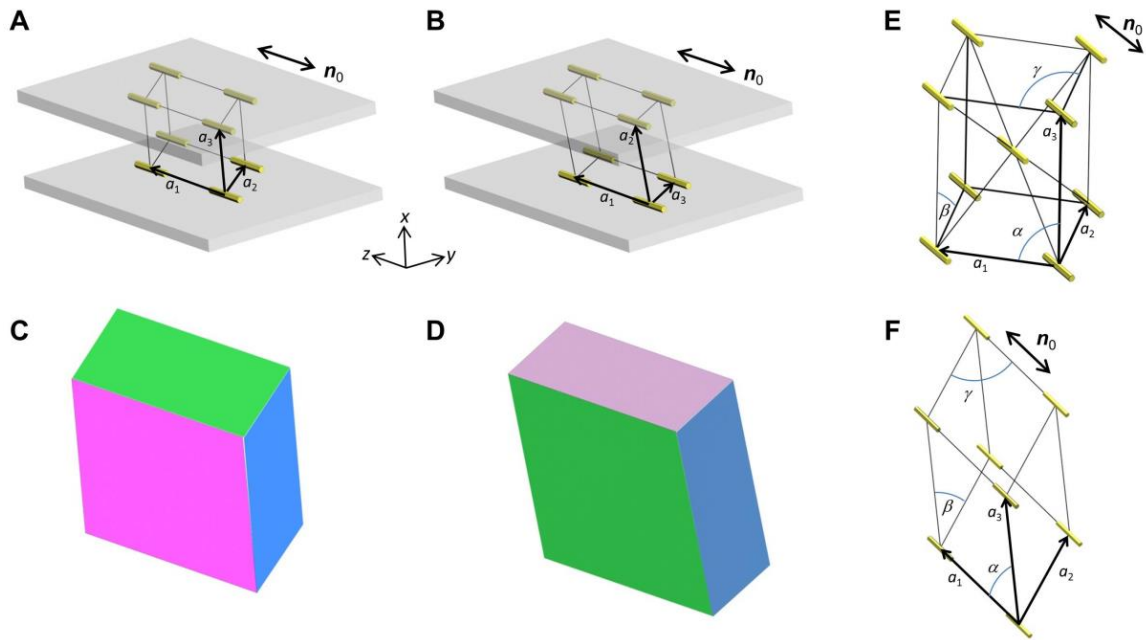

**Fig. S10. Tuning colloidal crystal symmetry through the control of material and cell parameters.** (A-D) Schematic representations of different orientations of the crystal lattice with respect to the confining cell substrates (A and C) for the orientation of the planes (100) and ($\bar{1}$00) parallel to substrates and (B and D) for the lattice rotated by 90° and with the planes (010) and ($0\bar{1}0$) parallel to substrates. (E) Schematic diagram of the potentially achievable body centered tetragonal crystalline organization of nanorods with θ = 45° and $a_1$=$a_2$≠$a_3$, α=β=γ=90°. (F) Potentially achievable trigonal rhombohedral lattice anticipated in the case of θ = 60°, with $a_1$=$a_2$=$a_3$ and 120°>α=β=γ≠90°.

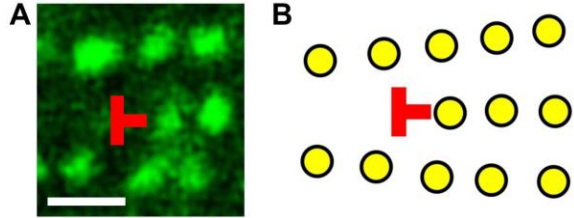

**Fig. S11. An edge dislocation in the triclinic colloidal crystal structure.** (A) Confocal luminescence in-plane image of nanorods in the triclinic crystal structure with edge dislocations. Scale bar, 2 μm. The image plane coincides with the crystallographic plane (100). (B) Schematic diagram of the edge dislocation shown in (A). The yellow filled circles depict centers of mass of nanorods within the colloidal crystal.

**Table S1.**
Experimentally measured parameters of a primitive lattice cell characterizing the triclinic colloidal crystal with the pinacoidal symmetry at different orientations of crystallographic planes. The top row corresponds to the data presented in Fig. S10A,C while the bottom row shows a triclinic lattice with a different orientation with respect to confining cell substrates (Fig. S10B,D). The values shown in the table are averaged over tens of measurements at different locations of the sample and an error represents the standard deviation.

| *Orientation* | $a_2/a_1$ | $a_3/a_1$ | α | β | γ |
|---|---|---|---|---|---|
| (100) upward | 0.66 ± 0.03 | 0.83 ± 0.05 | 56 ± 3 | 69 ± 3 | 49 ± 3 |
| (010) upward | 0.62 ± 0.02 | 0.67 ± 0.05 | 63 ± 4 | 44 ± 4 | 72 ± 4 |

# References

1. H. N. W. Lekkerkerker, R. Tuinier. *Colloids and the Depletion Interaction* (Springer Netherlands, 2011).
2. P. M. Chaikin, T. C. Lubensky, *Principles of Condensed Matter Physics* (Cambridge University Press, Cambridge, 1995).
3. V. N. Manoharan, Colloidal matter: Packing, geometry, and entropy. *Science* **349**, 1253751 (2015). doi:10.1126/science.1253751
4. P. F. Damasceno, M. Engel, S. C. Glotzer, Predictive self-assembly of polyhedra into complex structures. *Science* **337**, 453-457 (2012). doi:10.1126/science.1220869
5. B. Senyuk *et al.*, Topological colloids. *Nature* **493**, 200-205 (2013). doi:10.1038/nature11710
6. P. Poulin, H. Stark, T. C. Lubensky, D. A. Weitz, Novel colloidal interactions in anisotropic fluids. *Science* **275**, 1770-1773 (1997). doi:10.1126/science.275.5307.1770
7. S. Sacanna, W. T. M. Irvine, P. M. Chaikin, D. J. Pine, Lock and key colloids. *Nature* **464**, 575-578 (2010). doi: 10.1038/nature08906
8. M. R. Jones *et al.*, DNA-nanoparticle superlattices formed from anisotropic building blocks. *Nat. Mater.* **9**, 913-917 (2010). doi:10.1038/nmat2870
9. Q. Liu, Y. Yuan, I. I. Smalyukh, Electrically and optically tunable plasmonic guest-host liquid crystals with long-range ordered nanoparticles. *Nano Lett.* **14**, 4071-4077 (2014). doi:10.1021/nl501581y
10. B. Liu *et al.*, Switching plastic crystals of colloidal rods with electric fields. *Nat. Commun.* **5**, 3092 (2014). doi:10.1038/ncomms4092
11. S. K. Sainis, J. W. Merrill, E. R. Dufresne, Electrostatic interactions of colloidal particles at vanishing ionic strength. *Langmuir* **24**, 13334-13337 (2008). doi:10.1021/la8024606
12. A. Yethiraj, A. van Blaaderen, A colloidal model system with an interaction tunable from hard sphere to soft and dipolar. *Nature* **421**, 513–517 (2003). doi:10.1038/nature01328
13. M. F. Hsu, E. R. Dufresne, D. A. Weitz, Charge stabilization in nonpolar solvents. *Langmuir* **21**, 4881–4887 (2005). doi:10.1021/la046751m
14. T. C. Lubensky, D. Pettey, N. Currier, H. Stark, Topological defects and interactions in nematic emulsions. *Phys. Rev. E* **57**, 610–625 (1998). doi:10.1103/PhysRevE.57.610
15. R. W. Ruhwandl, E. M. Terentjev, Long-range forces and aggregation of colloid particles in a nematic liquid crystal. *Phys. Rev. E* **55**, 2958-2961 (1997). doi:10.1103/PhysRevE.55.2958
16. S. Ramaswamy, R. Nityananda, V. A. Raghunathan, J. Prost, Power-law forces between particles in a nematic. *Mol. Cryst. Liq. Cryst. Sci. Technol, Sect. A* **288**, 175-180 (1996). doi:10.1080/10587259608034594
17. C. P. Lapointe, T. G. Mason, I. I. Smalyukh, Shape-controlled colloidal interactions in nematic liquid crystals. *Science* **326**, 1083-1086 (2009). doi:10.1126/science.1176587
18. J. C. Loudet, P. Barois, P. Poulin, Colloidal ordering from phase separation in a liquid-crystalline continuous phase. *Nature* **407**, 611-613 (2000). doi:10.1038/35036539
19. A. Nych *et al.*, Assembly and control of 3D nematic dipolar colloidal crystals. *Nat. Commun.* **4**, 1489 (2013). doi:10.1038/ncomms2486
20. F. Wang *et al.*, Simultaneous phase and size control of upconversion nanocrystals through lanthanide doping. *Nature*, **463**, 1061-1065 (2010**)**. doi:10.1038/nature08777

21. H. Mundoor, I. I. Smalyukh, Mesostructured composite materials with electrically tunable upconverting properties. *Small* **11**, 5572-5580 (2015). doi:10.1002/smll.201501788
22. Materials and methods are available as supplementary materials on *Science* Online.
23. N. Bogdan, F. Vetrone, G. A. Ozin, J. A. Capobianco, Synthesis of ligand-free colloidally stable water dispersible brightly luminescent lanthanide-doped upconverting nanoparticles. *Nano Lett.* **11**, 835-840 (2011). doi:10.1021/nl1041929
24. D. E. Sands. *Introduction to Crystallography* (Dover Publications, 2012).
25. D. B. Conkey, R. P. Trivedi, S. R. P. Pavani, I. I. Smalyukh, R. Piestun, Three-dimensional parallel particle manipulation and tracking by integrating holographic optical tweezers and engineered point spread functions. *Opt. Express* **19**, 3835-3842 (2011). doi:10.1364/OE.19.003835
26. C. A. S. Batista, R. G. Larson, N. A. Kotov, Nonadditivity of nanoparticle interactions. *Science* **350**, 1242477 (2015). doi:10.1126/science.1242477
27. V. D. Nguyen, S. Faber, Z. Hu, G. H. Wegdam, P. Schall, Controlling colloidal phase transitions with critical Casimir forces. *Nat. Commun.* **4**, 1584 (2013). doi:10.1038/ncomms2597
28. R. W. Cahn, Materials science: Melting from within. *Nature* **413**, 582-583 (2001). doi:10.1038/35098169
29. R. C. Hayward, D. A. Saville, I. A. Aksay, Electrophoretic assembly of colloidal crystals with optically tunable micropatterns. *Nature* **404**, 56–59 (2000). doi:10.1038/35003530
30. I. I. Smalyukh, O. D. Lavrentovich, Anchoring-mediated interaction of edge dislocations with bounding surfaces in confined cholesteric liquid crystals. *Phys. Rev. Lett.* **90**, 085503 (2003). doi:10.1103/PhysRevLett.90.085503
31. J. C. Loudet, P. Hanusse, P. Poulin, Stokes drag on a sphere in a nematic liquid crystal. *Science* **306**, 1525 (2004). doi:10.1126/science.1102864
32. C. P. Royall, M. E. Leunissen, A. van Blaaderen, A new colloidal model system to study long-range interactions quantitatively in real space. *J. Phys. Condens. Matter.* **15**, S3581-S3596 (2003). doi:10.1088/0953-8984/15/48/017
33. G. J. Janz, S. S. Danyluk, Conductances of hydrogen halides in anhydrous polar organic solvents. *Chem. Rev.* **60**, 209-234 (1960). doi:10.1021/cr60204a005
34. J. N. Israelachvili. *Intermolecular and Surface Forces: With Applications to Colloidal and Biological Systems* (Academic Press, New York, 1985).
35. C. P. Royall, A. A. Louis, H. Tanaka, Measuring colloidal interactions with confocal microscopy. *J. Chem. Phys.* **127**, 044507 (2007). doi:10.1063/1.2755962
36. P. Sharma, A. Ward, T. Gibaud, M. F. Hagan, Z. Dogic, Hierarchical organization of chiral rafts in colloidal membranes. *Nature* **513**, 77-80 (2014). doi:10.1038/nature13694
37. L. M. Blinov, V. G. Chigrinov. *Electro optic Effects in Liquid Crystal Materials* (Springer-Verlag New York, Inc., New York, 1994).
38. S. H. Behrens, D. G. Grier, Pair interaction of charged colloidal spheres near a charged wall. *Phys. Rev. E* **64**, 050401(R) (2001). doi:10.1103/PhysRevE.64.050401
39. P. Schall, I. Cohen, D. A. Weitz, F. Spaepen, Visualization of dislocation dynamics in colloidal crystals. *Science* **305**, 1944-1948 (2004). doi:10.1126/science.1102186